\documentstyle[12pt]{article}

\catcode`\@=11
\long\def\@makefntext#1{ 
\protect\noindent \hbox to 3.2pt {\hskip-.9pt
$^{{\ninerm\@thefnmark}}$\hfil}#1\hfill} 

\def\thefootnote{\fnsymbol{footnote}}
 \def\@makefnmark{\hbox to 0pt{$^{\@thefnmark}$\hss}}  

\def\ps@myheadings{\let\@mkboth\@gobbletwo
\def\@oddhead{\hbox{} 
\rightmark\hfil\ninerm\thepage}
\def\@oddfoot{}\def\@evenhead{\ninerm\thepage\hfil 
\leftmark\hbox{}}\def\@evenfoot{}
\def\sectionmark##1{}\def\subsectionmark##1{}}

\begin{document}

\newcommand{\symbolfootnote}{\renewcommand{\thefootnote}
	{\fnsymbol{footnote}}}
\renewcommand{\thefootnote}{\fnsymbol{footnote}}
\newcommand{\alphfootnote}
	{\setcounter{footnote}{0}
	 \renewcommand{\thefootnote}{\sevenrm\alph{footnote}}}

\newcounter{sectionc}\newcounter{subsectionc}\newcounter{subsubsectionc}
\renewcommand{\section}[1] {\vspace{0.6cm}\addtocounter{sectionc}{1}
\setcounter{subsectionc}{0}\setcounter{subsubsectionc}{0}\noindent
	{\bf\thesectionc. #1}\par\vspace{0.4cm}}
\renewcommand{\subsection}[1] {\vspace{0.6cm}\addtocounter{subsectionc}{1}
	\setcounter{subsubsectionc}{0}\noindent
	{\it\thesectionc.\thesubsectionc. #1}\par\vspace{0.4cm}}
\renewcommand{\subsubsection}[1]
{\vspace{0.6cm}\addtocounter{subsubsectionc}{1}
	\noindent {\rm\thesectionc.\thesubsectionc.\thesubsubsectionc.
	#1}\par\vspace{0.4cm}}
\newcommand{\nonumsection}[1] {\vspace{0.6cm}\noindent{\bf #1}
	\par\vspace{0.4cm}}

\newcounter{appendixc}
\newcounter{subappendixc}[appendixc]
\newcounter{subsubappendixc}[subappendixc]
\renewcommand{\thesubappendixc}{\Alph{appendixc}.\arabic{subappendixc}}
\renewcommand{\thesubsubappendixc}
	{\Alph{appendixc}.\arabic{subappendixc}.\arabic{subsubappendixc}}

\renewcommand{\appendix}[1] {\vspace{0.6cm}
        \refstepcounter{appendixc}
        \setcounter{figure}{0}
        \setcounter{table}{0}
        \setcounter{equation}{0}
        \renewcommand{\thefigure}{\Alph{appendixc}.\arabic{figure}}
        \renewcommand{\thetable}{\Alph{appendixc}.\arabic{table}}
        \renewcommand{\theappendixc}{\Alph{appendixc}}
        \renewcommand{\theequation}{\Alph{appendixc}.\arabic{equation}}
        \noindent{\bf Appendix \theappendixc #1}\par\vspace{0.4cm}}
\newcommand{\subappendix}[1] {\vspace{0.6cm}
        \refstepcounter{subappendixc}
        \noindent{\bf Appendix \thesubappendixc. #1}\par\vspace{0.4cm}}
\newcommand{\subsubappendix}[1] {\vspace{0.6cm}
        \refstepcounter{subsubappendixc}
        \noindent{\it Appendix \thesubsubappendixc. #1}
	\par\vspace{0.4cm}}

\def\abstracts#1{{
	\centering{\begin{minipage}{30pc}\tenrm\baselineskip=12pt\noindent
	\centerline{\tenrm ABSTRACT}\vspace{0.3cm}
	\parindent=0pt #1
	\end{minipage} }\par}}

\newcommand{\bibit}{\it}
\newcommand{\bibbf}{\bf}
\renewenvironment{thebibliography}[1]
	{\begin{list}{\arabic{enumi}.}
	{\usecounter{enumi}\setlength{\parsep}{0pt}
\setlength{\leftmargin 1.25cm}{\rightmargin 0pt}
	 \setlength{\itemsep}{0pt} \settowidth
	{\labelwidth}{#1.}\sloppy}}{\end{list}}

\topsep=0in\parsep=0in\itemsep=0in
\parindent=1.5pc

\newcounter{itemlistc}
\newcounter{romanlistc}
\newcounter{alphlistc}
\newcounter{arabiclistc}
\newenvironment{itemlist}
    	{\setcounter{itemlistc}{0}
	 \begin{list}{$\bullet$}
	{\usecounter{itemlistc}
	 \setlength{\parsep}{0pt}
	 \setlength{\itemsep}{0pt}}}{\end{list}}

\newenvironment{romanlist}
	{\setcounter{romanlistc}{0}
	 \begin{list}{$($\roman{romanlistc}$)$}
	{\usecounter{romanlistc}
	 \setlength{\parsep}{0pt}
	 \setlength{\itemsep}{0pt}}}{\end{list}}

\newenvironment{alphlist}
	{\setcounter{alphlistc}{0}
	 \begin{list}{$($\alph{alphlistc}$)$}
	{\usecounter{alphlistc}
	 \setlength{\parsep}{0pt}
	 \setlength{\itemsep}{0pt}}}{\end{list}}

\newenvironment{arabiclist}
	{\setcounter{arabiclistc}{0}
	 \begin{list}{\arabic{arabiclistc}}
	{\usecounter{arabiclistc}
	 \setlength{\parsep}{0pt}
	 \setlength{\itemsep}{0pt}}}{\end{list}}

\newcommand{\fcaption}[1]{
        \refstepcounter{figure}
        \setbox\@tempboxa = \hbox{\tenrm Fig.~\thefigure. #1}
        \ifdim \wd\@tempboxa > 6in
           {\begin{center}
        \parbox{6in}{\tenrm\baselineskip=12pt Fig.~\thefigure. #1 }
            \end{center}}
        \else
             {\begin{center}
             {\tenrm Fig.~\thefigure. #1}
              \end{center}}
        \fi}

\newcommand{\tcaption}[1]{
        \refstepcounter{table}
        \setbox\@tempboxa = \hbox{\tenrm Table~\thetable. #1}
        \ifdim \wd\@tempboxa > 6in
           {\begin{center}
        \parbox{6in}{\tenrm\baselineskip=12pt Table~\thetable. #1 }
            \end{center}}
        \else
             {\begin{center}
             {\tenrm Table~\thetable. #1}
              \end{center}}
        \fi}

\def\@citex[#1]#2{\if@filesw\immediate\write\@auxout
	{\string\citation{#2}}\fi
\def\@citea{}\@cite{\@for\@citeb:=#2\do
	{\@citea\def\@citea{,}\@ifundefined
	{b@\@citeb}{{\bf ?}\@warning
	{Citation `\@citeb' on page \thepage \space undefined}}
	{\csname b@\@citeb\endcsname}}}{#1}}

\newif\if@cghi
\def\cite{\@cghitrue\@ifnextchar [{\@tempswatrue
	\@citex}{\@tempswafalse\@citex[]}}
\def\citelow{\@cghifalse\@ifnextchar [{\@tempswatrue
	\@citex}{\@tempswafalse\@citex[]}}
\def\@cite#1#2{{$\null^{#1}$\if@tempswa\typeout
	{IJCGA warning: optional citation argument
	ignored: `#2'} \fi}}
\newcommand{\citeup}{\cite}

\def\fnm#1{$^{\mbox{\scriptsize #1}}$}
\def\fnt#1#2{\footnotetext{\kern-.3em
	{$^{\mbox{\sevenrm #1}}$}{#2}}}

\font\twelvebf=cmbx10 scaled\magstep 1
\font\twelverm=cmr10 scaled\magstep 1
\font\twelveit=cmti10 scaled\magstep 1
\font\elevenbfit=cmbxti10 scaled\magstephalf
\font\elevenbf=cmbx10 scaled\magstephalf
\font\elevenrm=cmr10 scaled\magstephalf
\font\elevenit=cmti10 scaled\magstephalf
\font\bfit=cmbxti10
\font\tenbf=cmbx10
\font\tenrm=cmr10
\font\tenit=cmti10
\font\ninebf=cmbx9
\font\ninerm=cmr9
\font\nineit=cmti9
\font\eightbf=cmbx8
\font\eightrm=cmr8
\font\eightit=cmti8


\centerline{\tenbf THERMAL SELF-ENERGIES NEAR ZERO FOUR-MOMENTUM}
\vspace{0.8cm}
\centerline{\tenrm STAMATIS VOKOS}
\baselineskip=13pt
\centerline{\tenit Physics Department, FM-15, University of Washington}
\baselineskip=12pt
\centerline{\tenit Seattle, WA 98195, USA}
\vspace{0.9cm}
\abstracts{We demonstrate that one-loop self-energies at finite
temperature have a unique limit as the external four-momentum
goes to zero,
as long as the particles propagating in the loop have distinct masses.
We show that in spontaneously broken theories,
this result nonetheless does not affect the difference between
screening and propagating modes
and hence the usual resummed perturbation
expansion remains unaltered.
}

\vfil
\twelverm   
\baselineskip=14pt
\section{Introduction and Motivation}
\vspace*{-0.3cm}
This talk is a review of work that I have done with P.~Arnold,
P.~Bedaque, and A.~Das \cite{abdv}.
In finite temperature field theory, the existence of an additional
four-vector, namely the four-velocity of the plasma, allows one to
construct two independent Lorentz scalars on which all Green's functions,
and in particular, polarization
tensors and self-energies can depend, namely
$\omega = P\cdot u\,$ and
$k=\left([(P\cdot u)^2-P^2]\right)^{1\over 2}\,$.
Here $u^\mu$ is the four-velocity of the plasma and $P^\mu=(p^0,\vec{p})$ is
the four-momentum of any particle.
In the rest-frame of the heat bath, these scalars reduce to $p^0$ and
$p=|\vec{p}|$ respectively.

This separate dependence allows one
to take the limits $p^0\rightarrow 0$ and $p\rightarrow 0$ in
different orders. In general, one
expects that the limits need not commute, since they correspond to
different physical situations. For instance, one may imagine computing
the change in the free energy of the QED plasma, after placing
two {\em static} charges $q_1$ at $\vec{r}_1$ and $q_2$ at
$\vec{r}_2$, as a function of their separation $r=|\vec{r}_1-\vec{r}_2|$.
Linear response theory gives the answer
\begin{equation}
U(r) =q_1q_2 \int {d^3\vec{k}\over (2\pi)^3} e^{i\vec{k}\cdot
\vec{r}}\, {1\over \vec{k}^2+\Pi_{00}(0,|\vec{k}|)}\,.
\end{equation}
For large separations, the integral is dominated by $\vec{k}\approx
0$. So, one may effectively replace
\begin{equation}
\Pi_{00}(0,k)\to \lim_{k\to 0}\Pi_{00}(0,k)\equiv \lim_{k\to 0}
\lim_{k^0\to 0}\Pi_{00}(k^0,|\vec{k}|)\,.
\end{equation}
Denoting this double limit by $m_{e\ell}^2$, the square of the
electric screening mass
of the photon, one obtains the usual expression for $r\to \infty$,
\begin{equation}
U(r)\to {q_1q_2\over 4\pi} {e^{-m_{e\ell}r}\over r}\,.
\end{equation}

My interest in understanding the structure of thermal self-energies
near zero four-momentum was motivated by the prospect of baryogenesis
at the electroweak phase
transition.  However, electroweak baryogenesis has brought the need to
understand the detailed dynamics of
the transition\cite{ewpt}. It is well-known that the validity of perturbation
theory at finite
temperature is seriously compromised by infrared divergences.
Therefore, one needs to resum (infinite) sets of diagrams in order to
improve convergence. This means, among other things, using
``self-consistent'' propagators, which entails replacing tree-level
masses by their higher-order values.
To achieve that goal, one needs to solve self-consistently the
computed dispersion relations
\begin{equation}
P^2=m^2+\Pi(P^2)\,,
\end{equation}
for the location of the physical pole (to
that order), and then use that value in an improved propagator.
That is where the behavior of thermal self-energies enters with a vengeance.
For instance, it has been shown that
in the case of hot QCD\cite{therm},
self-interacting scalars\cite{analytic,weldondisc}, and gauge theories
with chiral fermions\cite{ferm},
the two aforementioned limits of the self-energy do not indeed commute.
Guided by these results, people have been using the non-analytic (in
$p^0$ and $p$)
high-temperature expressions in the improved propagators in the
Standard Model.
In this talk, however, I will
demonstrate that there exist contributions to the one-loop self-energy of a
massive gauge boson in a
spontaneously broken gauge theory, which possess a {\em unique} limit
as $p$ and
$p^0$ tend to zero, as long as the particles propagating in the loop
have different masses. Given that the Standard Model is such a theory,
does that invalidate the literature results of carefully computed quantities at
the phase transition? The answer is {\em no}. I will show, that
even if one-loop self-energies are perfectly analytic ``around
zero four-momentum'',
the usual approximation which uses the non-commuting limits is the relevant and
correct procedure, at least for the purposes of computing {\it
physical} quantities, such as poles of particle propagators.

\section{Spontaneously Broken U(1) Theory}
\vspace{-.3cm}

For simplicity, we will perform the calculation of the polarization
tensor for the massive vector boson in the Abelian Higgs model in
unitary gauge. Unitary gauge is
infamous for complications in the Higgs sector at
finite temperature\cite{puzzle}. In the gauge sector, however, these
complications are absent and the smaller number of diagrams makes its
use preferable for our purposes.

The Lagrangian for the Abelian Higgs model in the
unitary gauge is given by
\begin{eqnarray}
{\cal L}&=&-{1\over 4}F^{\mu\nu} F_{\mu\nu} + {e^2 v^2\over 2} A^\mu
A_\mu
+ {1\over 2} \partial^\mu\eta\partial_\mu\eta - {m^2\over 2} \eta ^2\nonumber\\
&\phantom{=}&+{e^2\over 2}A^\mu A_\mu\eta^2+e^2 v A^\mu A_\mu\eta
-\lambda v \eta^3
-{\lambda\over 4} \eta^4 ,
\end{eqnarray}
\noindent
where $\eta$ is the Higgs field, $A_\mu$ is the U(1) gauge field and
the vacuum expectation value, $v=m/\sqrt{2\lambda}$.
In unitary gauge, there is a single one-loop, momentum-dependent
correction to the photon propagator,
which we denote by
$\tilde\Pi_{\mu\nu}$. This diagram gives via the usual methods\cite{kapusta},
\begin{eqnarray}
{\rm Re}\tilde\Pi_{00}^\beta=4e^2\int {d^3\vec{k}\over (2\pi)^3}
&\!\!\!\Big[&\!\!\!\!
{n(\omega_k)\over 2\omega_k}\,{M^2-(p_0+\omega_k)^2\over
(p_0+\omega_{k})^2-\Omega_{k+p}^2}
+{n(\Omega_k)\over 2\Omega_k}\,{M^2-\Omega_k^2\over
(p_0+\Omega_k)^2-\omega_{k+p}^2}\Big]\nonumber\\
+&\!\!\!(&\!\!\!\!p_0\to -p_0)\,.
\end{eqnarray}
Here we have defined $M=ev$, $\omega_k=\sqrt{\vec k^2+m^2}$ and
$\Omega_k=\sqrt{\vec k^2+M^2}$.
After doing the angular integration, one obtains
\begin{eqnarray}
{\rm Re}\tilde\Pi_{00}^\beta(p_0,p)=-{e^2\over 2\pi^2}\int_0^\infty dk\, k
&\!\!\!\!\Bigg[&\!\!\!\!
{(k^2+p_0^2+\Delta)n(\omega_k)\over2\omega_k}{1\over p}
\ln\left|S_1\right|+{k^2n(\Omega_k)\over2\Omega_k}{1\over p}\ln\left|S_2
\right|\nonumber\\
&\!\!\!\!+&\!\!\!\!n(\omega_k){p_0\over p}
\ln\left| {(p_0^2-p^2+\Delta)^2-4(p_0\omega_k+pk)^2\over
(p_0^2-p^2+\Delta)^2-4(p_0\omega_k-pk)^2}\right|\Bigg]\,,
\end{eqnarray}
where $\Delta = m^2-M^2$, and the $S_i$ are given by the following
expression, with $m_1=m$ and
$m_2=M$,
\begin{equation}
S_i={(p_0^2-p^2+2pk-(-1)^i \Delta)^2-4p_0^2\omega_i^2\over
(p_0^2-p^2-2pk-(-1)^i \Delta)^2-4p_0^2\omega_i^2}\,,\qquad\quad i=1,2.
\end{equation}

Let us analyze the small-$p^0$, small-$p$ behavior of Eq.\ (7).
For that purpose, let us set
\begin{equation}
p^0=\alpha p\,.
\end{equation}
Then, for {\it nonzero}\ values of $\Delta =m^2-M^2$, it is
clear that
\begin{equation}
\lim_{p\to 0}{\rm Re}\tilde\Pi_{00}^\beta(\alpha p,p)=
-{4e^2\over \pi^2}\int_0^\infty dk\left[k^2{n(\omega_k)\over 2\omega_k}+
{k^4\over m^2-M^2}\left({n(\omega_k)\over2\omega_k}-{n(\Omega_k)\over
2\Omega_k}\right)\right]\,.
\end{equation}
In particular, this limit is finite, $\alpha$-independent and hence
independent of the ratio $p_0/p$ as $p_0$ and $p$ approach zero.
Alternatively, this may be obtained by simply putting $P^\mu=0$ in
Eq.\ (6). So, the double limit is unique, as promised.
Furthermore, it is easy to establish that ${\rm Re}\tilde\Pi_{ii}^\beta$
has a unique double limit as well.

The high-temperature limit of Eq.\ (10) can be easily obtained to be
\begin{equation}
\lim_{{p\to 0}\atop {p_0\to 0}}{\rm
Re}\tilde\Pi_{00}^\beta(p_0,p)={1\over 6}e^2 T^2\,.
\end{equation}
This turns out to be the same as the $(p_0=0,\vec{p}\to 0)$ limit of
the equal mass case $\Delta =0$ (see also Eq.\ (13)). Note that
even though Eq.\ (10) appears to be singular when
$m=M$, it indeed has a finite limit as the two masses become
degenerate and corresponds to the $p_0=0, \vec{p}\to 0$ limit of the
degenerate case.

\pagebreak[4]

\section{Debye and Plasmon Masses}
\vspace{-.3cm}
At this point it is instructive to compare and contrast the above
result with the usual non-commuting double limits. This exercise will
shed light on what one means by considering the self-energy ``near
zero external four-momentum''.
Let's start with Eq.\ (6).
We Taylor expand
the denominators of the integrand in the high-temperature limit
$T\gg m_i,p_0,p$,
keeping in mind that $k\sim T$ (in view of the Bose-Einstein factor).
For instance,
\begin{eqnarray}
{1\over(p_0+\omega_{k})^2-\Omega_{k+p}^2}&=&
{1\over 2 P\cdot K+(P^2+m^2-M^2)}\nonumber\\
&=&{1\over 2 P\cdot K}\left(1-{P^2+m^2-M^2\over 2 P\cdot K}+
\ldots\right)\,,
\end{eqnarray}
where $K^0$ is on the mass-shell. Then one finds
that all masses drop out from the integrand (or can be neglected to leading
order). Therefore, it is not surprising to find that the
high-temperature limit in {\em this} regime of external momenta is
\begin{equation}
\tilde\Pi^\beta_{00}(p_0,p)= {e^2T^2\over 6} \int
{d\Omega_n\over 4\pi}
\left[ {n_0^2P^2\over (n\cdot P)^2}-{2n_0p_0\over n\cdot P}\right]\,,
\end{equation}
which agrees with the standard Braaten-Pisarski result\cite{therm,puzzle},
and which is explicitly non-analytic.
Here $n^\mu=(1,\vec{n})$, with $|\vec{n}|=1$, and the angular
integration is over all possible orientations of that vector.

So, where is the sleight of hand? The same expression, Eq.\ (6), surely cannot
be simultaneously analytic and non-analytic around zero. The answer lies
in the study of the validity of the Taylor expansion above. The
non-analytic expression was got by assuming that $P\cdot K \gg
\Delta$, or $p_0,p\gg |m^2-M^2|/T$. For a theory with $\Delta = 0$,
this is always satisfied. However, with $\Delta\neq 0$, there is a
region
$p_0,p\ll |m^2-M^2|/T$, for which the Taylor
expansion above is inappropriate and the analysis of
the previous section shows that the self-energy has a unique value
around zero.

\section{Discussion}
\vspace{-.3cm}

It is easy to see that the same result holds for a theory with two
scalars, as well as for QED with massive fermions: the
one-loop self-energy/polarization tensor at finite temperature
has a unique limit as the external four-momentum goes to zero.
The absence of the usual non-commuting double limits is traced to the fact
that there is (generically) a finite mass difference among the particles
propagating in the loop. One can understand this result in the following
way. The real part of the one-loop self-energy is related to the imaginary
part through the dispersion relation\cite{weldonfp},
\begin{eqnarray}
{\rm Re}\Sigma^\beta_R(p_0,p)&=&{1\over \pi}{\cal P}\int_{-\infty}^\infty
du {{\rm Im}\Sigma^\beta_R(u,p)\over u-p_0}\nonumber\\
&=&{2\over\pi}{\cal P}\int_0^\infty du\,u{{\rm Im}\Sigma^\beta_R(u,p)\over
u^2-p_0^2}\,.
\end{eqnarray}
The last equality follows from the fact that ${\rm Im}
\Sigma^\beta_R(p_0,p)$
is an odd function of $p_0$\cite{jeon}. Here $\Sigma^\beta_R$ is the
retarded two point function related to $\Sigma^\beta$ by
\begin{eqnarray}
{\rm Im}\Sigma^\beta_R(u,p)&=&{\rm Im}\Sigma^\beta(u,p)\,{\rm tanh}{\beta
u\over 2}\nonumber\\
{\rm Re}\Sigma^\beta_R(u,p)&=&{\rm Re}\Sigma^\beta(u,p)\, .
\end{eqnarray}

As pointed out by
Weldon\cite{weldondisc}, ${\rm Im}\Sigma^\beta_R(u,p)$ is
non-zero only for some values of $u^2-p^2$. The imaginary part of the
self-energy is expressed in terms of the discontinuity of $\Sigma^\beta_R
(p_0,p)$ along these cuts on the real axis,
\begin{equation}
\lim_{\epsilon\to 0^+}\left(\Sigma^\beta_R(p_0+i\epsilon,p)
-\Sigma^\beta_R(p_0-i
\epsilon,p)\right)=-2i{\rm Im}\Sigma^\beta_R(p_0,p)\,,
\end{equation}
for real $p_0$. For fixed $m_1$ and $m_2$, these cuts exist for
\begin{eqnarray}
u^2 -p^2&\geq& (m_1+m_2)^2\,, \\
u^2 -p^2&\leq& (m_1-m_2)^2\,.
\end{eqnarray}
The first cut is the usual zero-temperature cut corresponding to
the decay of the incoming particle, whereas the second
appears only at $T\neq 0$ and represents absorption of a particle from
the medium.
The first cut does not lend itself to non-commuting double limits, so
the only suspect is the second cut. In fact, it is this cut which is
responsible for the non-commuting double limits in the case
$m_1=m_2$\cite{weldonfp}. In our case however, the contribution
of this cut is perfectly well-behaved as $P^\mu\to 0$.
In fact, if we denote this contribution by $C_2(p_0,p)$, then we obtain
\begin{equation}
{\rm Re}\Sigma^\beta_R(p_0,p)\ni C_2(p_0,p)={2\over\pi}
{\cal P}\int_0^{\left(p^2+(m_1-m_2)^2\right)^{1\over 2}}
 du\,u {{\rm Im}\Sigma^\beta_R(u,p)\over u^2-p_0^2}\,.
\end{equation}
Performing the change of variables $u\rightarrow
u/\sqrt{p^2+(m_1-m_2)^2}$, we obtain
\begin{equation}
{\rm Re}\Sigma^\beta_R(p_0,p)\ni C_2(p_0,p)={2\over\pi}
{\cal P}\int_0^1
 du\,u {{\rm Im}\Sigma^\beta_R(u\sqrt{p^2+(m_1-m_2)^2},p)
\over u^2-{p_0^2\over p^2+(m_1-m_2)^2}}\,.
\end{equation}
As long as the masses are different, the zero momentum
limit of $C_2(p_0,p)$ is well-defined and given by
\begin{equation}
C_2(0,0)={2\over\pi}
\int_0^{|m_1-m_2|} du {{\rm Im} \Sigma^\beta_R(u,0)\over u}\,.
\end{equation}
This limit, however, is not well-defined if the masses are equal.
Note that Eq.\ (21) is well-behaved, given that
${\rm Im}\Sigma^\beta(u,0)$
is odd in $u$, and goes as $u$ for small $u$.

One may naturally wonder whether our observation has any effect on
standard computations of {\it physical} quantities, such as the
difference between Debye and plasmon masses in the standard electroweak
theory, and whether there could be any effect on studies of the
electroweak phase transition. In fact it does not, as
one may argue in view of the results of the previous section.
There, we noted that
our result for the $P^\mu\to 0$ limit, Eq.\ (11), depends on assuming
$p_0,p\ll |\Delta|/T$ in Eq.\ (6), since Eq.\ (6) is dominated
by $k\sim T$. However, the region of interest
for self-consistently finding the Debye or plasmon poles of the vector
propagator is when $p_0$ or $p$ take values of order
$m_i\gg |\Delta|/T$.
In that regime, $\Delta$ can be
ignored in Eq.\ (7), in which case one recovers the usual non-commuting
double limits. For
$p_0$ and $p$ small compared to $|\Delta|/T$, the functions
$\Pi^\beta_{00}(p_0,0)$ and $\Pi^\beta_{00}(0,p)$ tend to the same limit.
However, at order $m$, the functions take on different values. As the mass
difference goes to zero, it is clear that the unique limit disappears,
as well.

\end{document}